\documentclass[showpacs,aps,prb,amsmath,amssymb,superscriptaddress]{revtex4}
\usepackage{graphicx}
\usepackage{dcolumn}
\usepackage{bm}

\begin{document}

\title{Strains and pseudo-magnetic fields in circular graphene rings}

\author{Nima Abedpour}
\address {School of Physics, Institute for Research in
Fundamental Sciences, IPM, Tehran 19395-5531, Iran}
\author{Reza Asgari}
\email{asgari@ipm.ir}
\address {School of Physics, Institute for Research in
Fundamental Sciences, IPM, Tehran 19395-5531, Iran}
\author{F. Guinea}
\address {Instituto de Ciencia de Materiales de Madrid, CSIC, Sor Juana In\'es de la Cruz 3, E-28049 Madrid, Spin}

\begin{abstract}
We demonstrate that circular graphene ring under a shear stress
displays strong pseudo-magnetic fields. We calculate the
pseudo-magnetic field both from continuum elasticity theory as well as
molecular dynamics simulations. Stable wrinkles are induced by shear deformations and lead to enhancement of the pseudo-magnetic field. The strong
pseudo-magnetic field found here can be observed by imaging
graphene flake at the atomic level {\it e.g.} through scanning tunneling
microscope.
\end{abstract}


\maketitle

\section{introduction}

Graphene has recently attracted intensive interest as a promising
candidate material for the new generation of electronics and
spintronics~\cite{geim}. One of the exciting physics on graphene is
strain exerted on graphene samples ~\cite{vozmediano,pereira,guinea1}. It was
proposed that strain can be utilized to generate various basic
elements for all-graphene electronics~\cite{pereira}.

Semiconductor quantum rings have been investigated by a number of
groups~\cite{bayer,ribeiro,Retal08,YPSR010,Hetal10}. In graphene ring, the spectrum reveals
signatures of effective time-reversal symmetry breaking, in which
the spectra are most naturally interpreted in terms of effective
magnetic flux contained in the ring, even when no real flux is
present. Quantum rings can be considered as prototypical devices in
mesoscopic physics, as they show one of the most basic coherence
effects, namely the Aharonov-Bohm effect, oscillations of the
transmission as a function of the magnetic flux through the ring.
The reason for these oscillations is the phase difference between
electrons traveling along the different arms of the ring.
Furthermore, Benjamin and Pachos~\cite{benjamin} proposed creating a
ring of single-layer graphene in which they induce d-wave
superconductivity via the proximity effect or directly make graphene
superconducting by doping. The quantum qubits would be built around
the $\pi$-junction that naturally occurs in graphene and would not
require bilayer structures. The Aharonov –Bohm oscillations, on the
other hand, has been observed~\cite{Hetal10} in a graphene ring, consisting of a planar honeycomb lattice of carbon atoms in a ring shape, by
changing the voltage applied to the side gate or the back gate.

When the graphene sheet is under tension, the side contacts
induce a long-range elastic deformation which acts as a
pseudo-magnetic field for its massless charge
carriers~\cite{suzuura, maes}. This is because strain changes the bonds length between atoms and affects the way electrons move among them.
 The pseudo-magnetic field would reveal itself through its effects on electron orbits. The tension can be generated either by the
electrostatic force of the underlying gate~\cite{fogler} by interaction
of graphene with the side walls~\cite{bunch}, as a result of thermal
expansion~\cite{bao} or by quench height fluctuations~\cite{guinea}. A particular strain geometry in graphene could
lead to a uniform pseudo-magnetic field and might open up interesting applications in graphene nano-electronics
with real magnetic fields~\cite{low}. Mechanical strains can introduce new environments in
studying novel physics of graphene.

It is commonly believed that strains have important influence on
the electronic structure of graphene~\cite{neto}. A graphene ring is a particularly convenient geometry. Strains can possibly be
manipulated efficiently in samples with good adhesion to the
substrate, such as graphene layers grown epitaxially on {\rm SiC}.
Recently, the physical properties of graphene when its hexagonal
lattice is stretched out of equilibrium have been investigated by
many groups~\cite{exp, teague}. Scanning tunneling microscopy
studies on graphene surface have indeed revealed a correlation
between local strain and tunneling conductance~\cite{teague}.
Motivated by experiments pointing to a remarkable stability of
graphene with large strains, we have carried out theoretical analysis and the molecular
dynamics simulation to explore the
pseudo-magnetic field in strained graphene ring.

In this papers we focus on a particular aspect of the physics of
graphene ring namely the appearance of gauge fields and its
corresponding pseudo-magnetic fields which arise in the
circular graphene ring when a shear force is applied to its
boundary. The common belief is that the morphologies of graphene
surface under strain are negligible. The aim of this paper is to
show that, contrary to these expectations, the structure of a
deformed surface can lead to a strong pseudo-magnetic field.

The paper is organized as follows. In Sec. II we introduce our
model and formalism. In Sec. III, our numerical results for the
strain and the pseudo-magnetic field for deformed graphene ring are
presented. Finally, we conclude in Sec. IV with a brief summary.

\section{Model and Theory}

We analyze both analytically and numerically the strains and pseudo-magnetic fields in circular graphene ring
under a shear stress. A shear force is applied to the boundary,
inducing shear deformations inside graphene ring.

The properties of graphene presented lie on the special character
of its low energy excitations obeying a two dimensional
massless Dirac equation. The graphene ring with valley degree of freedom,
$\tau=\pm 1$ for the inequivalent $K$ and $K'$ valleys, is
modelled by the massless Dirac Hamiltonian in the continuum
model~\cite{slon,haldane}
\begin{center}
${\cal H}_0=\hbar v_{\rm F} \tau \left(\sigma_1 \, k_1 + \sigma_2 \,
k_2 \right)$
\end{center}
where $k_i$ is an envelope function
momentum operator, $v_{\rm F}\simeq 10^6 m/s$ is the Fermi velocity,
and $\sigma_i=x,y,z$ are the Pauli matrices that act on the sublattice
pseudospin degree of freedom. Hence it is
important to establish the robustness of the low energy description under small lattice deformations.
The concepts of gauge fields and covariant derivatives can be translated into the language of differential geometry,
based on differential forms. When ideal graphene is distorted,
the effective Hamiltonian will be changed into
\begin{eqnarray}
H=v_{\rm F}~(~\bf{p}-e~\bf{A})\cdot \bf{\sigma}
\end{eqnarray}
where the induced vector potential field is defined through the deformations of
sample and $\bf {p}$ is the momentum in the polar coordinate. The induced
gauge fields can be calculated through the following
expressions~\cite{suzuura}

\begin{eqnarray}\label{Amiu}
v_{\rm F} e A_{x}&=& \hbar g_2 (u_{xx}-u_{yy})\\ \nonumber v_{\rm F} e
A_{y}&=& - 2 \hbar g_2 u_{xy}
\end{eqnarray}

with

\begin{eqnarray}
g_2 &=&\frac{3\kappa \beta}{4}~t\\ \nonumber \kappa
&=&\frac{\sqrt{2}\mu}{2 B}
\end{eqnarray}
where $t\approx 2.7$ eV is the nearest-neighbor hopping parameter
and $\beta=\partial \ln(t)/\partial \ln(a)\simeq 2$ is the electron
Gr\"{u}neisen parameter. For the shear $\mu$ and bulk moduli, $B$ we
have used~\cite{zakharchenko} the values $\mu=9.95$ eV $\AA^{-2}$ and $B =12.52$ eV
$\AA^{-2}$. We thus find that $\kappa\approx 0.56$. Once the induced
gauge filed is obtained, the pseudo-magnetic field can be
calculated by $\bm{B}= \nabla\times\bm{A}$.

When the graphene is deformed due to the force exerts on
the boundaries, the strain tensor can be calculated from
\begin{eqnarray}
u_{\alpha\beta}=\frac{\partial_{\alpha}u_{\beta}+
\partial_{\beta}u_{\alpha}}{2}+\frac{\partial_{\alpha}h\partial_{\beta}h}{2}
\end{eqnarray}
where $\bm{u}$ is the atomic displacement field and $h$ is the out of
plane displacement. The contribution of the out-of-plane atomic
displacements are noticeable in our numerical calculations.

\subsection{Analytical expression for pseudo-magnetic filed}

We consider the case of a graphene ring of radius $R$ and width $W
= R-R_1$ where the clamped outer and inner boundaries are circles
of radius $R$ and $R_1$, respectively. Additionally, we assume that the
out-of-plane atomic displacements being zero, $h=0$. The
displacement at the outer boundary is $u_\theta (R , \theta ) =
U_\theta , u_r ( R , \theta ) = 0$, which the shear deformation at
the boundary is defined by $U_\theta$, and at the inner boundary
we have $u_\theta (R_1 , \theta ) = u_r ( R_1 , \theta ) = 0$. The
displacements in the graphene ring are
\begin{align}
u_r ( r , \theta ) &= 0 \nonumber \\
u_\theta ( r , \theta ) &= U_\theta \left[ -\frac{R_1^2 R}{r ( R^2 - R_1^2 )}+\frac{R r}{R^2-R_1^2}
 \right]~.
\end{align}
The second term in the expression for $u_\theta ( r , \theta )$ is a
pure rotation, required to satisfy the boundary conditions. The only
non-zero component of the strain tensor is $u_{r \theta}$. The
strain at radius $r$ is
\begin{equation}
u_{r \theta} ( r , \theta ) = \frac{1}{2} \left( \frac{\partial
u_\theta}{\partial r} + \frac{1}{r} \frac{\partial u_r}{\partial
\theta} - \frac{u_\theta}{r} \right) = \frac{ U_\theta R_1^2 R}{r^2 (
R^2 - R_1^2 )}
\end{equation}
The maximum strain at the boundary is actually given by $\bar{u} = U_\theta
 R/(R^2-R_1^2)$. Using polar coordinates $(r,\theta)$, Eq.~(\ref{Amiu}) can be rewritten as

\begin{eqnarray}\label{analy}
A_r = &\Phi_0&\frac{c \beta}{ a} [(\frac{\partial u_r}{\partial r}-\frac{u_r}{r}-\frac{1}{r}\frac{\partial u_{\theta}}{ \partial \theta})\cos( 3\theta)\nonumber\\
&+&(-\frac{\partial u_{\theta}}{\partial r}+\frac{u_{\theta}}{r}-\frac{1}{r}\frac{\partial u_{\theta}}{ \partial \theta})\sin( 3\theta)]~, \nonumber\\
A_\theta =&\Phi_0&\frac{c \beta}{ a}[(-\frac{\partial u_{\theta}}{\partial r}+\frac{u_{\theta}}{r}-\frac{1}{r}\frac{\partial u_{r}}{ \partial \theta})\cos( 3\theta)\nonumber\\
&+&(-\frac{\partial u_{r}}{\partial r}+\frac{u_{r}}{r}+\frac{1}{r}\frac{\partial u_{\theta}}{ \partial \theta})\sin( 3\theta)] ~,
\end{eqnarray}
where $\Phi_0=h/2 e$ is the quantum unit of magnetic flux and $c=\sqrt{3}\kappa/(2\pi)$ is a constant. Furthermore, the induced gauge field in the polar coordinate is given by

\begin{eqnarray}\label{analy}
A_r =\Phi_0\frac{c \beta U_{\theta}}{ a}\frac{2 R_1^2R}{r^2(R^2-R_1^2)}\sin(3\theta) \nonumber\\
A_\theta =\Phi_0\frac{c \beta U_{\theta}}{ a}\frac{2 R_1^2R}{r^2(R^2-R_1^2)}\cos(3\theta)
\end{eqnarray}

Eventually, the pseudo-magnetic field acting on the electrons is simply given by
\begin{equation}\label{beff}
B ( r , \theta ) = \Phi_0 \frac{c \beta}{ a} \frac{8 U_\theta
R_1^2 R}{r^3(R^2-R_1^2)} \cos ( 3 \theta )
\end{equation}

It would be worthwhile mentioning that the pseudo-magnetic field diverges near the clamped sites as $B ( r) \approx 8 c \Phi_0 \beta
\bar{u} R_1^2 / ( a r^3 )$.

\subsection{Molecular dynamics simulation}

We used Molecular Dynamics simulation (MD) to simulate a suspended
circular graphene ring, consists of a planar honeycomb lattice of
carbon atoms in a ring shape. Rigidly clamped boundary conditions
were employed. We simulated the system at different temperatures
by employing Nos\`{e}-Hoover thermostat to help the system reach
equilibrium at a given temperature. Our present results are limited
to $T = 50 K$. In this work, we used both the Brenner's bond-order potential ~\cite{brenner,
khodemoon} incorporating the second nearest neighbors interactions through their dependence on the bond angles and the second-generation reactive empirical bond-order
(REBO) potential~\cite{rebo} for the carbon-carbon interactions (for more details see Appendix A). It is well established that
MD simulations predict the correct mechanical properties and the form of
the structures with carbon atoms by using the Brenner potential.

The number of carbon atoms in our simulations is $5125$ corresponding to
graphene ring of radius $R = 7$ nm and width $W = 4$ nm. At the
beginning of calculations, we simulate the circular graphene ring
at a given temperature with circular clamped boundary conditions.
After reaching a stable configuration, we rotate the outer
boundary of graphene ring about $\theta_0 = 12^{\circ}$ as the
system shown in Fig.~\ref{sample}( left panel). For later
purposes, we name the system as case one. In this case, the
displacement of the boundary atoms is about $U_{\theta} =
R\pi\theta_0/180$. For the sake of comparison, we would like to consider another system in which the inner boundary is pulled down. In
the later case, in order to avoid from strong wrinkles, we reduce
the rotation of the outer boundary about $6^{\circ}$ and then pull
slowly down the central part of graphene sheet about $z=-1.7$nm as
shown in Fig.~\ref{sample}( right panel). Again, the
latter system is labeled as case two.

\section{Numerical Results}

From the analytical calculations, the pseudo-magnetic field obtained in the graphene ring geometry
where $h=0$ is shown in Fig.~\ref{pin_0}. For given the circular radius, $B=7.89\times 10^7 \cos( 3\theta)/{r_0}^3$
in units of tesla where $r_0$ scaled in units of Angstrom. The maximum field occurs in the vicinity of the inner boundary
and it decreases like $1/r^3$. Although the analytical model described above can explain the behavior of the strains
and magnetic field,  we  point out that the wrinkle structures represent an important feature of the results.

We use two different systems in the numerical simulations. We first study a
graphene ring which is placed in the $x-y$ plane and set at a given
temperature by allowing the surface fluctuations in the
third dimension, and then the outer boundary of graphene ring is
rotated while the inner boundary is clamped as is shown in Fig.~1 (
left panel). In case one, we have seen noticeable wrinkle
structures around the inner boundary along the perpendicular
direction (z-axis). In case two, on the other hand, the graphene ring surface is smoother, but the average carbon-carbon distance is longer than the one obtained in case one.
In the following, we will describe two cases and find the
pseudo-magnetic fields emerge due to the strains. In our MD
simulations, the equations of motion are integrated by the Verlet
algorithm with a time step of $0.5$ fs. Our numerical results show that the formation of the wrinkles are quantitatively sensitive to the details of the potential. We simulate a system with two mentioned potentials and find that the order of the pseudo-magnetic fields are the same ( for more details, see Appendix A ). Therefore, we mainly use the Brenner potential~\cite{brenner} to calculate the pseudo-magnetic filed.

\subsection{Deformed structure}

\begin{figure}[h]
\begin{center}
\includegraphics[width=3.7in]{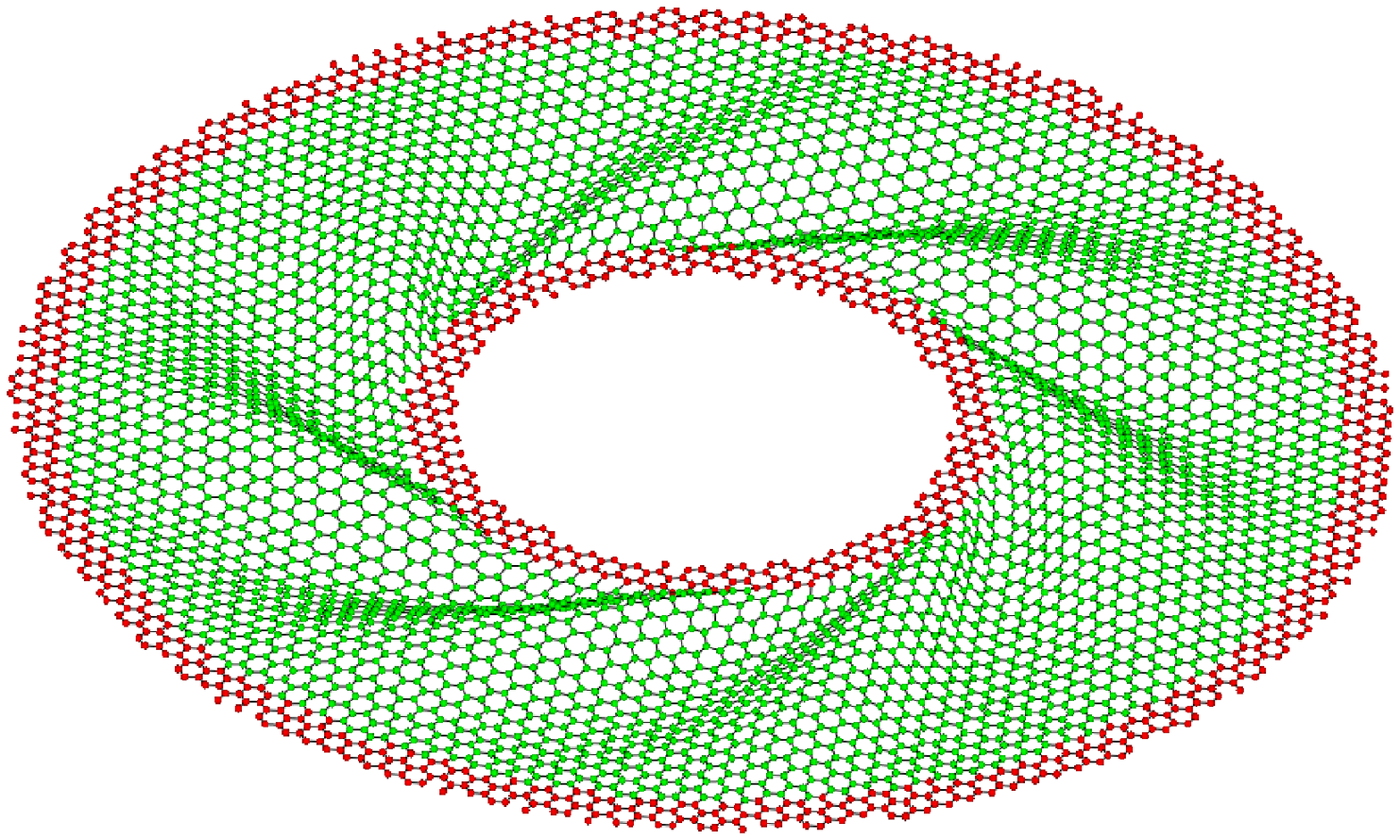}
\includegraphics[width=3.3in]{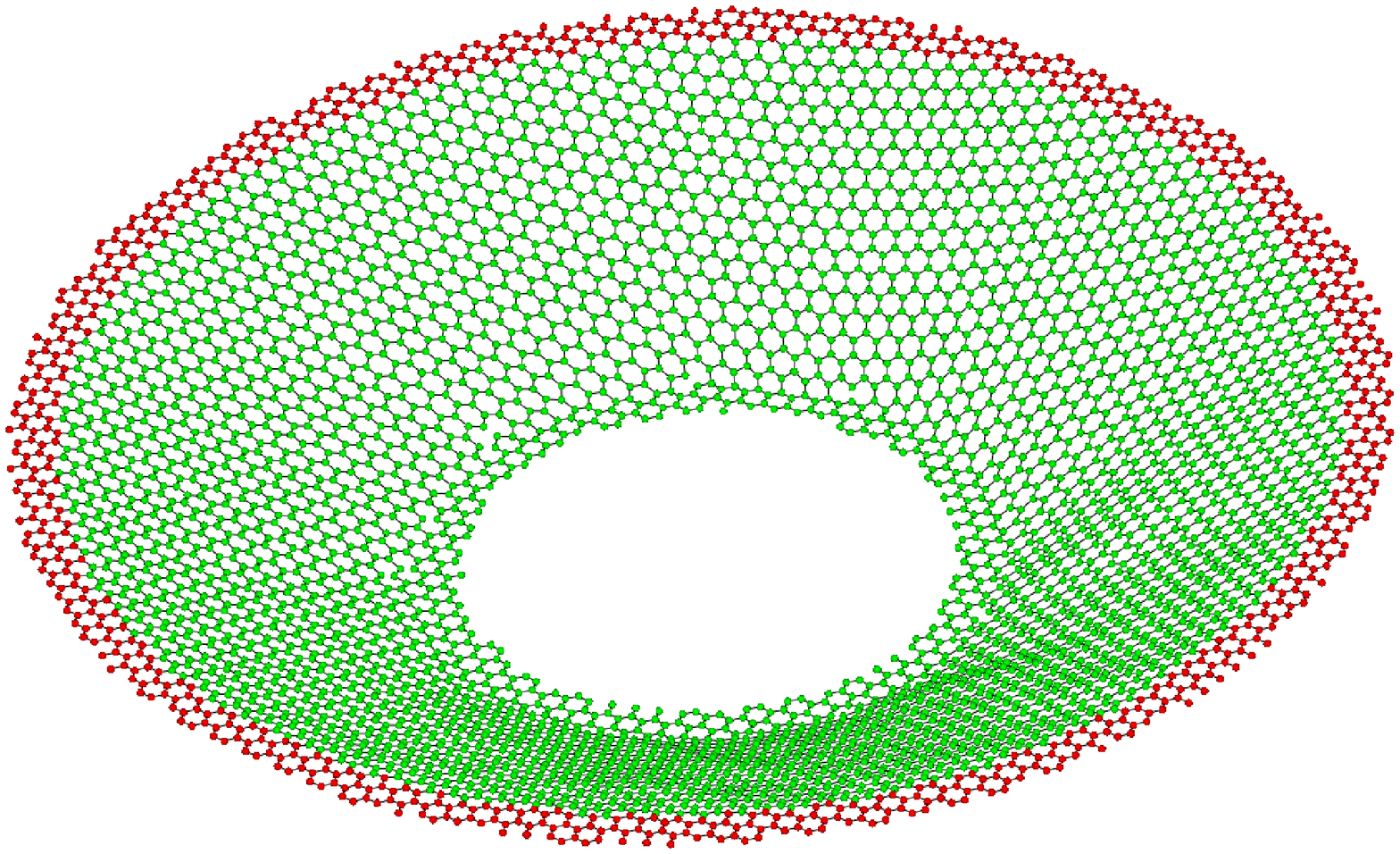}
\caption{( Color online) A representative atomic configuration in
MD simulations for circular graphene ring at  $T=50$ K. Left panel:
case one in which the rotation of the inner boundary is
$12^{\circ}$ in $x-y$ plane. Right panel, case two in which the
inner boundary has been rotated about $6^{\circ}$ and afterwards
the central part has been pulled down about $z=-1.7$ nm
.}\label{sample}
\end{center}
\end{figure}

As it can be seen from the left panel of Fig~\ref{sample}, the
main part of the deformation occurs in the area near the inner
boundary and makes noticeable fluctuations along the
$z$-direction. Due to the appearance of wrinkles in the system,
the strain increases near the inner boundary and reduces in the
far regions. In contrast to the analytical calculations, the strain does
not behave as $r^{-2}$. The analytical model does not describe well a system with out of plane wrinkles.

In the right panel of Fig~\ref{sample}, we show case two, where the static wrinkles are approximately washed out and seemingly the surface of graphene ring is
much smoother than case one. The structure of surface
displacements can also be seen in Fig~\ref{u}. This figure clearly shows that
the morphology of the surfaces are enormously different and we expect to have a stronger pseudo-magnetic fields for case one with respect to case two.

It should be noticed that the structure of the simulated samples
weakly depend on temperature up to $T=300$K.

\subsection{Pseudo-magnetic field}

Since the structure of wrinkles, after the relaxation of
graphene's shape, is static, we  can calculate the gauge fields and also the
pseudo-magnetic fields from Eq.~(\ref{Amiu}).

Our numerical results
for the induced gauge field calculated from the strains are illustrated in Fig.~\ref{amiu}
for two mentioned cases in the $x-y$ plane. We take a time average over the atom positions in order to pass over the thermal fluctuation effects and study only the wrinkling structures. The length value of the vector denotes the
absolute value of the gauge field. For case one, the induced gauge fields are not uniform in the area which wrinkles appear.

The pseudo-magnetic field can be calculated from
the induced gauge field and its distributions are demonstrated in
Fig.~\ref{bz}. It is worth mentioning that the non-zero pseudo-magnetic
field, $B_z$ almost occurs along the wrinkles, however its
value and structure depends on each wrinkle structure. This
is due to the fact that the structure of the wrinkles is different
along different lattice directions such as armchair and zigzag. The value of the pseudo-magnetic field that increases with increasing curvature of the wrinkles mostly occur
around the central part.

The structure of the pseudo-magnetic field along one of
wrinkles, which is indicated in the inset figure, is illustrated in Fig~\ref{bzsd} (left panel) for case one. Notice that $B_z$ decreases
for $x<0$ along the wrinkle. The curve can be fitted quite well with the expression $\exp(-\alpha x)$ where $\alpha=1.2 \pm 0.1$,
 which displays fast decaying of the pseudo-magnetic field close to the outer boundary. The pseudo-magnetic field behaves randomly for $x>0$.
 The latter behavior is due to the fact that the path is no longer along a specific wrinkle.
 As it can be observed from the results, the pseudo-magnetic field value is immense. Similar massive pseudo-magnetic fields raised by highly strained
nanobubbles that form when graphene is grown on a platinum surface have been measured by Levy {\it et al.}~\cite{levy}.

For case two, which wrinkles are
disappeared, the symmetry of the graphene is clearly induced in the $B_z$,
(see Fig~\ref{bz}, right panel ).

In order to have a better understanding regarding the structure of
the pseudo-magnetic fields, we calculate the pseudo-magnetic
fields for two cases along the azimuthal angel at given radius,
$r$. Fig.~\ref{bzcircle}, shows the pseudo-magnetic field $B_z$ as
a function of $\theta$ around a ring for radius $r=4$ nm in case
one (left panel) and moreover the pseudo-magnetic field is shown
at $r=6.5$ nm for case two (right panel). In the latter case, we
have seen the behavior most likely as $B_z \propto \cos(
3\theta)$. In this figure, we have plotted function, which is
proportional to $ \cos( 3\theta)$ as solid lines. Apparently the
result in the left panel refers to case one behaves similarly to
case two with respect to the azimuthal variable, however there are
detectable fluctuations due to the appearance of the wrinkles.

The analytical expression of the pseudo-magnetic field acting
 on the electrons is $B(r,\theta)\propto \frac{1}{r^3}\cos( 3\theta)$ when atomic out-of-plane displacements are ignored.
The expression was obtained for a system that lies in the $x-y$ plane, $h=0$, where there are no wrinkle structures on the flake.
Since case two, which has less wrinkle structures, is somehow similar to the aforementioned system, we have found numerically the
$\cos( 3\theta)$ dependence for the pseudo-magnetic field. However due to the fact that the inner boundary is pulled down,
the pseudo-magnetic field behaviors differently as a function of $r$.

\section{Conclusions}

In summary, we have investigated the strains and pseudo-magnetic
fields in circular graphene ring under a shear stress. We find,
from the elastic theory, the induced gauge filed as function of the maximum strain at
the boundary, $\bar{u}$. The magnitude of the pseudo-magnetic
field near the boundary is $B \approx 4 \sqrt{3} \Phi_0 \beta \kappa \bar{u}
/ ( \pi a R )$. Moreover, The field diverges near the clamped sites as
$B ( r ) \propto 1 / r^3$, where $r$ is the distance to the site
while the strains diverge as $1 / r^2$. From numerical simulation results, we find
wrinkles structures on graphene flakes and the pseudo-magnetic field is the same order of magnitude that obtained by analytical calculations. We also find that the wrinkle structures represent an important feature of the pseudo-magnetic field.
In addition, it is also shown that
the pseudo-magnetic field behaves like as $\cos ( 3 \theta )$. These
results are essential for understanding the electronic properties
of graphene ring and its strain engineering for potential
applications.

\section{ACKNOWLEDGMENTS}
We thank A. Fognini and A. Naji for their useful comments. This research was supported in part by the Project of Knowledge Innovation Program (PKIP) of Chinese Academy of Sciences, Grant No. KJCX2.YW.W10.
F. G. is supported by  by MICINN (Spain), grants FIS2008-00124 and CONSOLIDER CSD2007-00010.

\appendix

\section{ Empirical potentials }

We used both the Brenner's bond-order potential ~\cite{brenner} incorporating the second nearest neighbors interactions through their dependence on the bond angles and the second-generation reactive empirical bond-order
(REBO) potential~\cite{rebo} for the carbon-carbon interactions. The latter potential is based on the empirical bond-order formalism and allows for covalent
bond binding and breaking with associated changes in the
atomic hybridization. Consequently, such a classical potential
allows to model complex chemistry in large many-atom
systems. The Brenner bond-order potential can be written in
the following general form for the binding energy,
\begin{equation}
E_b=\sum_i~\sum_{j>i}~ \{ V^R(r_{ij})-b_{ij} V^A(r_{ij})\}
\end{equation}

The first term is repulsive, and the second one is attractive.
$r_{ij}$ is the distance between pairs of nearest-neighbor atoms $i$
and $j$. Although this expression is a simple sum over bond
energies, it is not a pair potential since the $b_{ij}$, which is called
the bond-order factor, is in essence a many-body factor. The many-body nature of
$b_{ij}$ makes the bond energy depend on the local environment
of the bond. This feature allows the Brenner potential to
predict correct geometries and energies for many different
carbon structures. The empirical bond-order function used here
is written as a sum of terms $ b_{ij}=[b_{ij}^{\sigma-\pi}+b_{ji}^{\sigma-\pi}]/2+b_{ij}^{\pi}$ where values for the functions $b_{ij}^{\sigma-\pi}$ and $b_{ji}^{\sigma-\pi}$ depend on the local coordination and bond angles for
atoms $i$ and $j$ , respectively. The first term is a function of the bond angles similar to that in the Brenner potential~\cite{brenner} while the second term incorporates the third nearest neighbors via a bond-order term associated with the dihedral angles and becomes nonzero upon bending of the graphene sheets. The values for all the parameters used in our calculation for the potentials can be found in Refs.~{\onlinecite{brenner, rebo}} and are therefore not listed here.

We used the Brenner potential~\cite{khodemoon} and alternatively the REBO potential by using Large-scale Atomic or Molecular Massively Parallel Simulator package~\cite{lammps} ( LAMMPS) for carrying out the molecular dynamic simulations in this work. We calculated the pseudo-magnetic
fields for the two different potentials along the azimuthal angel at given radius,
$r\simeq 40$\AA~. Fig.~\ref{bzpot}, shows the pseudo-magnetic fields $B_z$ as
a function of $\theta$ along a circular ring of radius $r=4$ nm. Note that the order of the pseudo-magnetic field is the same in both cases and moreover, we see the same envelop behavior at long-wavelength region. However, they exhibit different oscillations modes at short wavelength due to the way in which the bond-order ( the dihedral angle) is handled in the REBO potential.

\section{Simulation methods}

Computer simulations generates very detailed information at the microscopic level and the conversion of this information into macroscopic level is the province of statistical mechanics. Molecular dynamics, on the other hand, is an important tool to investigate the microscopic
behaviors by integrating the motions of particles or particle clusters. In the molecular dynamics, the trajectories of atoms are determined by numerically solving Newton's equations of motion for a many-body interacting systems, where forces between the particles and their potential energy are defined by certain force fields. We used NVT ensemble where the system is isolated from changes in moles (N), volume (V) and temperature (T). In NVT, the energy of endothermic and exothermic processes is exchanged with a thermostat. A variety of thermostat methods is available to add and remove energy from the boundaries of a MD system in a nearly  realistic way, approximating the canonical ensemble. On the other hand, a micro-canonical molecular dynamics trajectory may be seen as an exchange of potential and kinetic energy, with total energy being conserved. For every time step, each particle's position and velocity may be integrated with a method such as Verlet. Given the initial positions and velocities, all future positions and velocities can be calculated. If there is a large enough number of atoms, statistical temperature can be estimated from the instantaneous temperature, which is found by equating the kinetic energy of the system to $nk_BT/2$ where $n$ is the number of degrees of freedom of the system.

We simulated the system at nonzero temperatures by employing a Nos\'{e}-Hoover thermostat and the time step is taken as $0.5$ fs. At the beginning of the simulation, we consider a circular graphene ring incorporating atoms which are located in the $x-y$ plane. The initial structures are firstly optimized giving the carbon-carbon bond length of $1.45$ \AA~and all samples are initially relaxed at a desired temperature for a duration $1.5$ ns. In atomic positions at the boundaries, we enforce $z=0$ to prevent atomic motion along the $z$ direction and furthermore, after some simulation run-times, we set $x=y=$const. to clamp atoms.

\newpage

\begin{figure}
\begin{center}
\includegraphics[width=4.in]{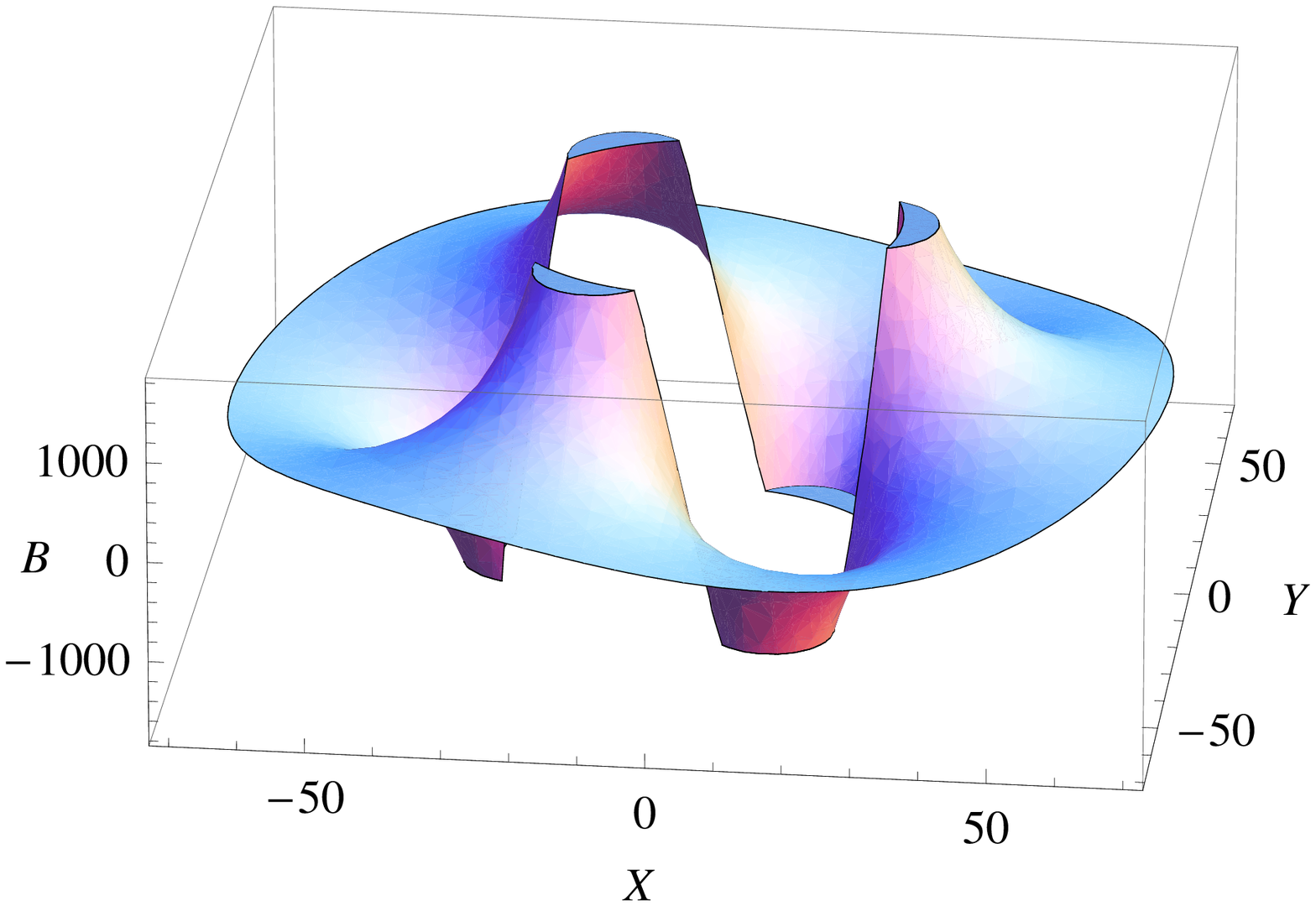}
\caption[fig]{(Color online) Pseudo-magnetic field, given by Eq.~\ref{beff} in units of tesla induced by shear
strains in a circular graphene ring. Here, $R=70$, $R_1=30$\AA and $\theta=12^{\circ}$. } \label{pin_0}
\end{center}
\end{figure}

\begin{figure}[h]
\begin{center}
\includegraphics[width=3.3in]{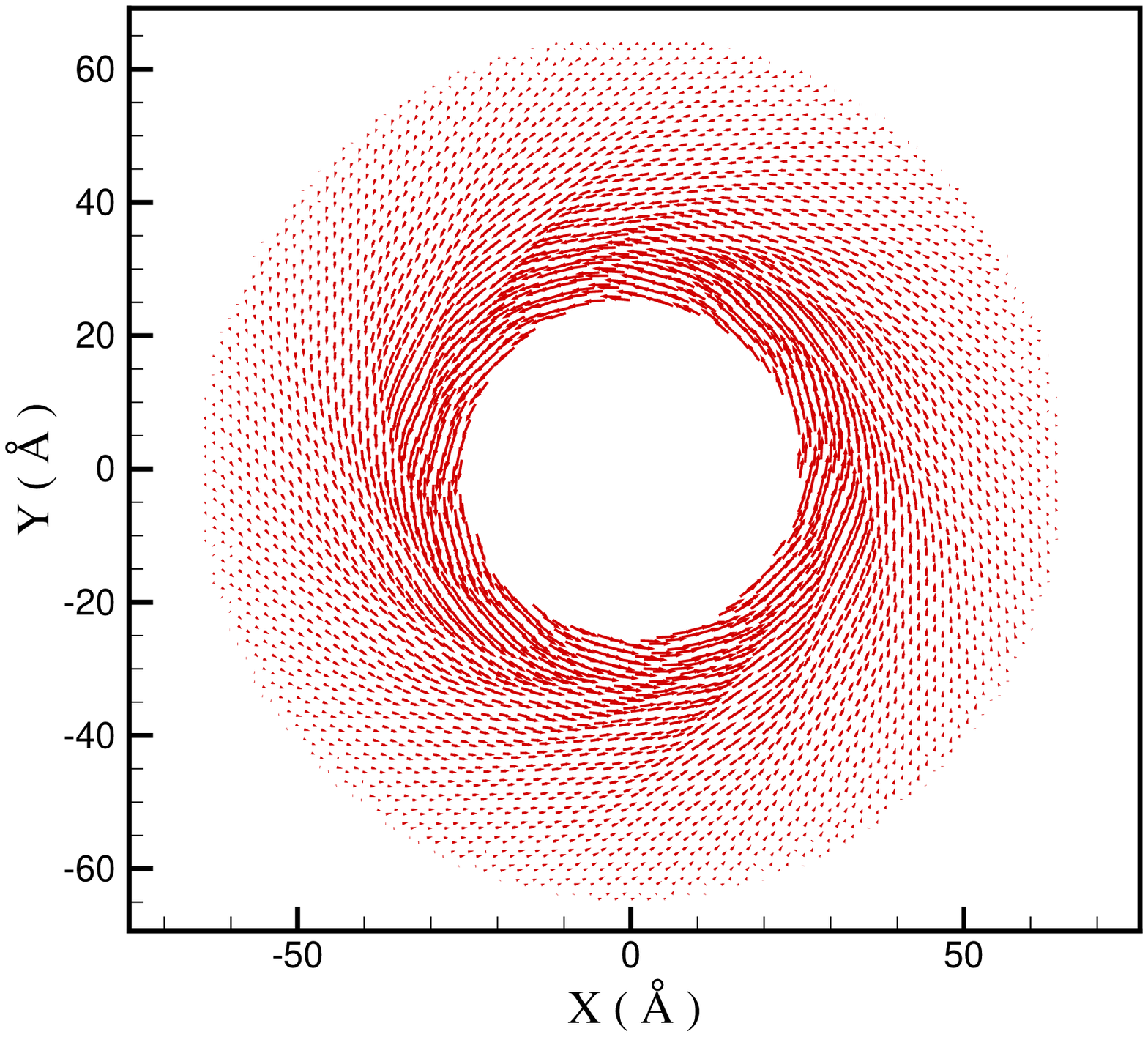}
\includegraphics[width=3.3in]{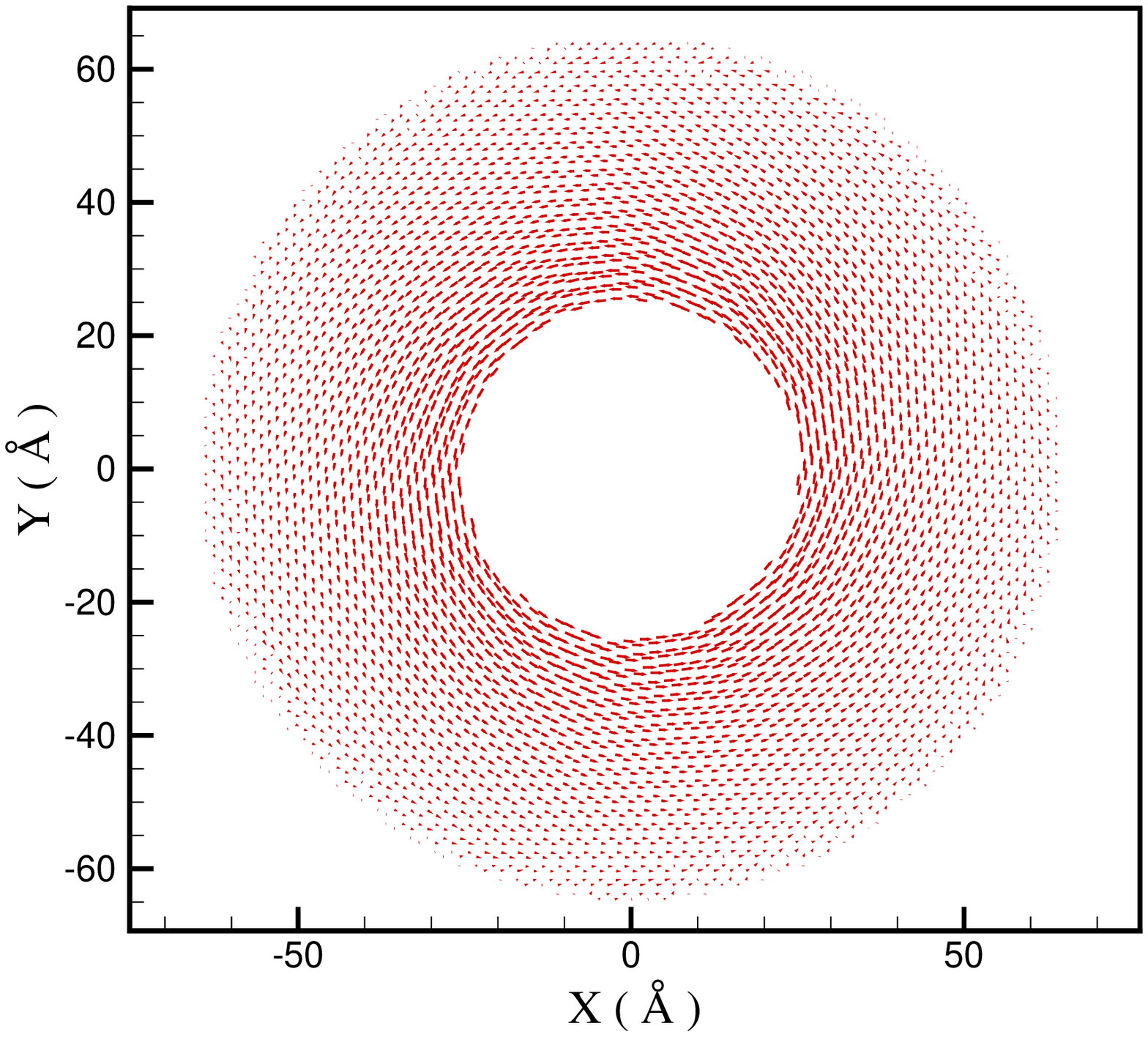}
\caption{(Color online) Projected atomic displacements for the
case one, left panel and case two, right panel.}\label{u}
\end{center}
\end{figure}

\begin{figure}[h]
\begin{center}
\includegraphics[width=3.3in]{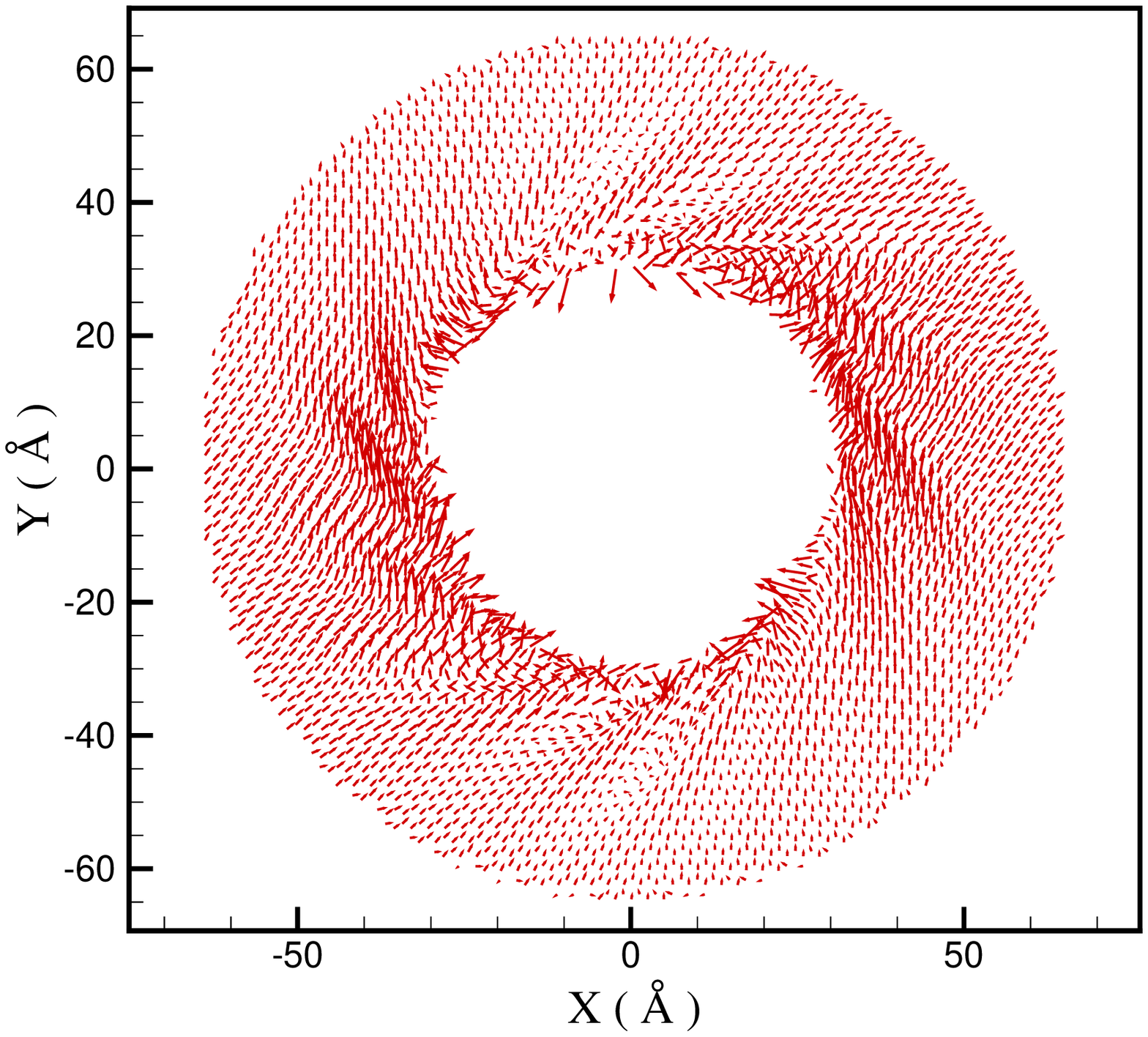}
\includegraphics[width=3.3in]{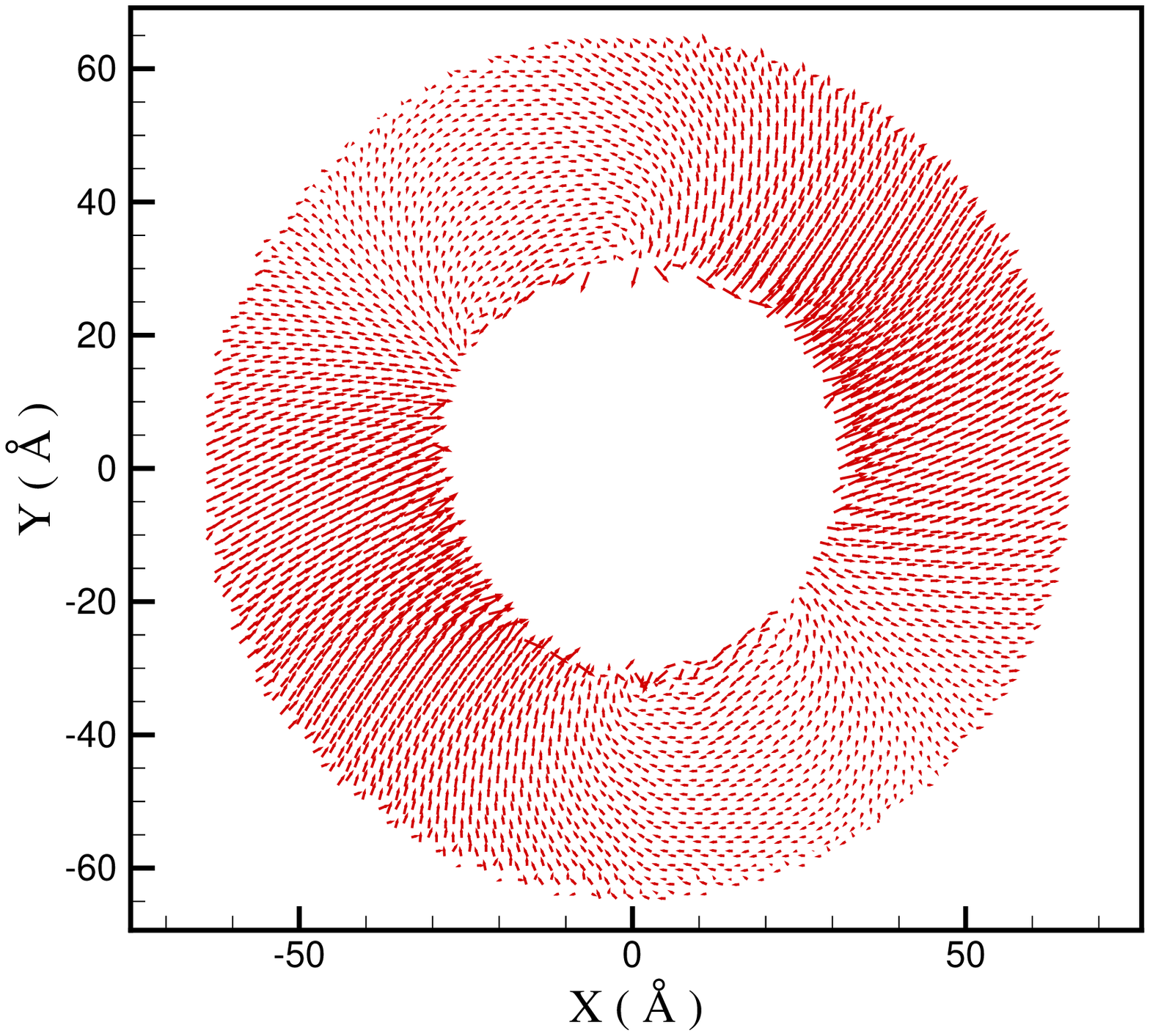}
\caption{(Color online) Projected the gauge filed, $A_{\mu}$ for two mentioned
cases. Notice that the length value of the vector denotes the
absolute value of the gauge filed. }\label{amiu}
\end{center}
\end{figure}

\begin{figure}[h]
\begin{center}
\includegraphics[width=3.4in]{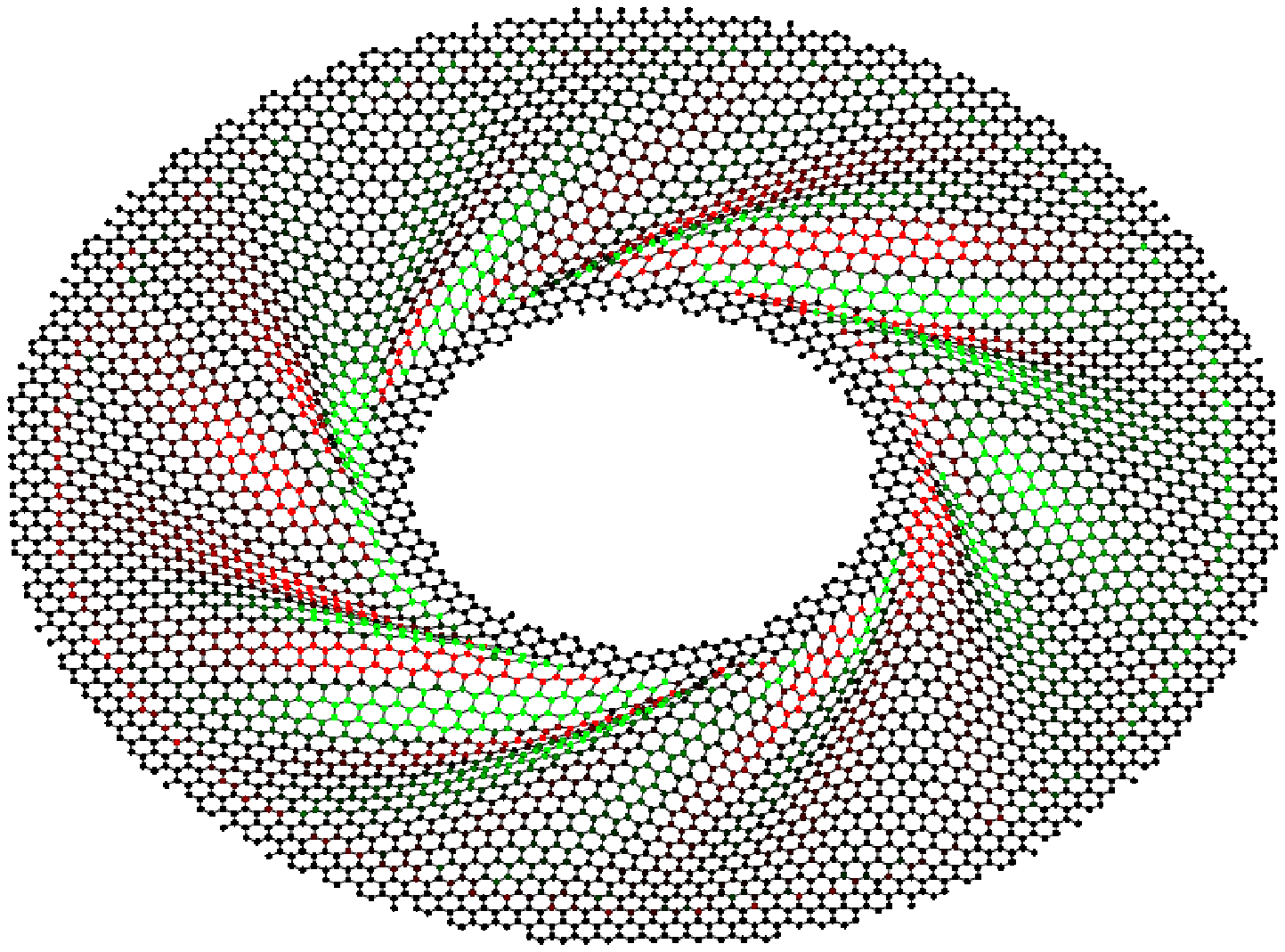}
\includegraphics[width=3.4in]{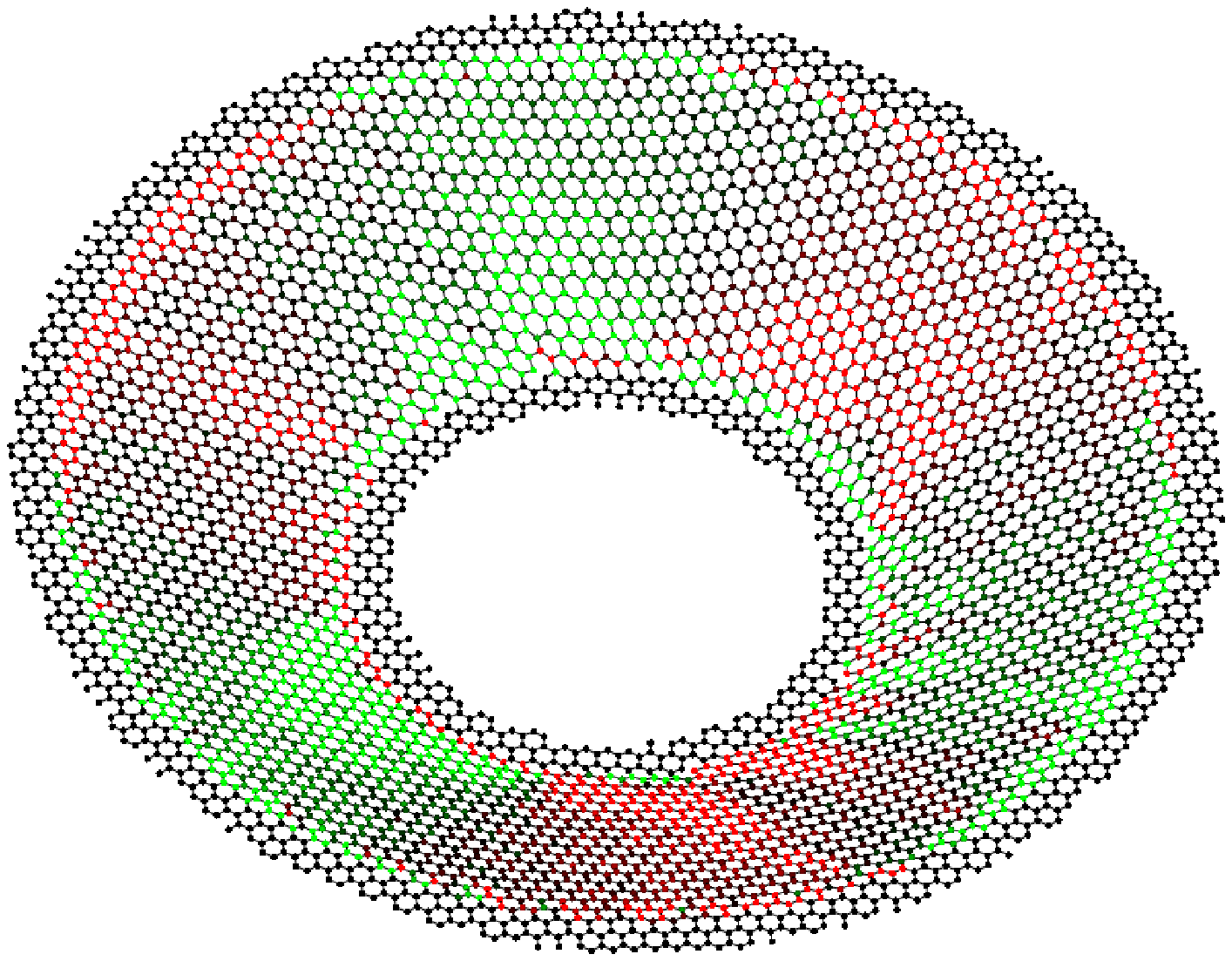}
\caption{(Color online) Distribution of the pseudo-magnetic field,
$B_z$ on graphene ring for two mentioned cases. Green, red and
black colors correspond to positive, negative and zero values of
the pseudo-magnetic field, respectively. For the case one, the
dark-green(or dark-red) correspond to the strongest
pseudo-magnetic filed which is about $1000 T$ and for the case
case two, it is about $200 T$}\label{bz}
\end{center}
\end{figure}

\begin{figure}[h]
\begin{center}
\includegraphics[width=3.3in]{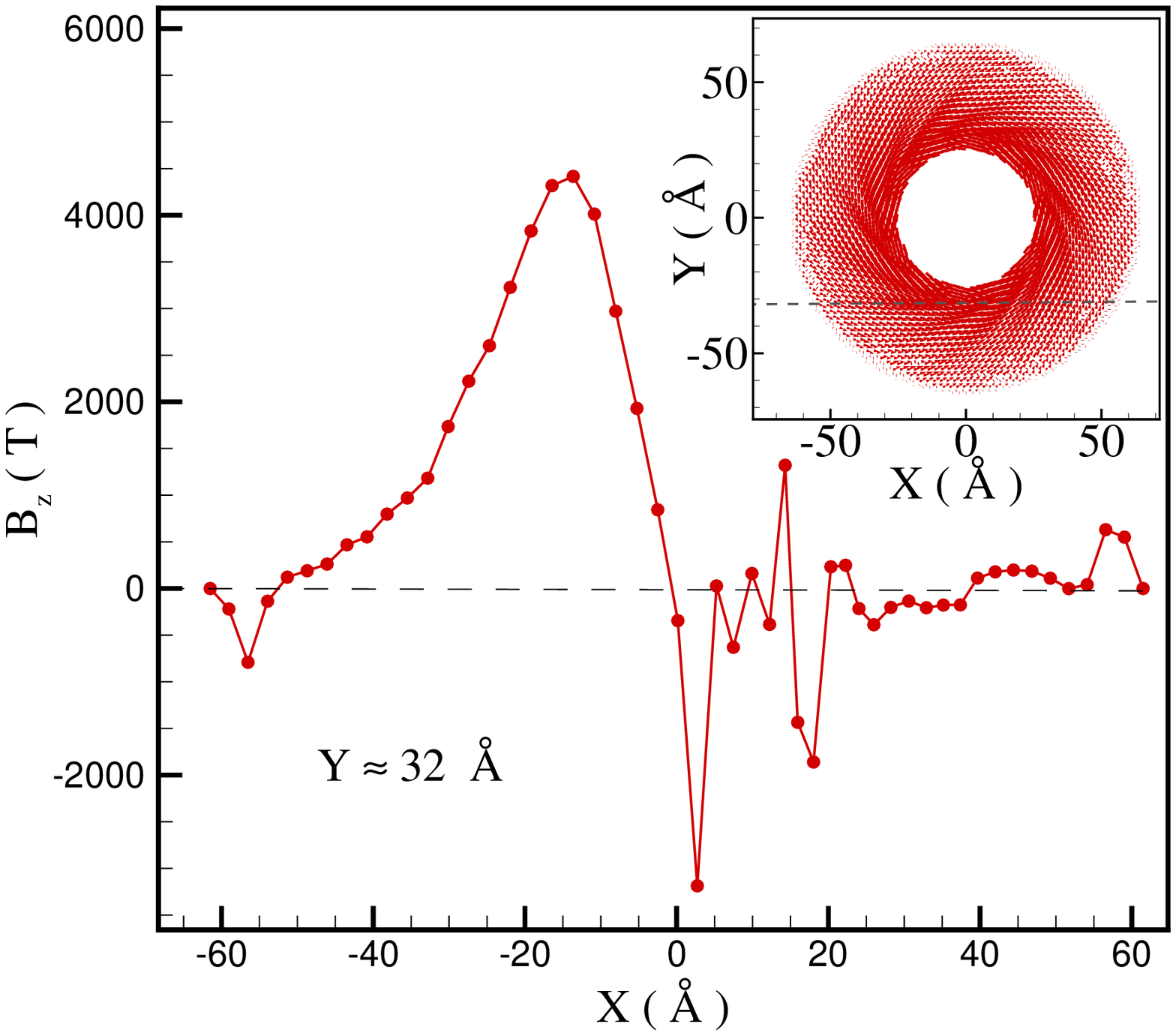}
\includegraphics[width=3.3in]{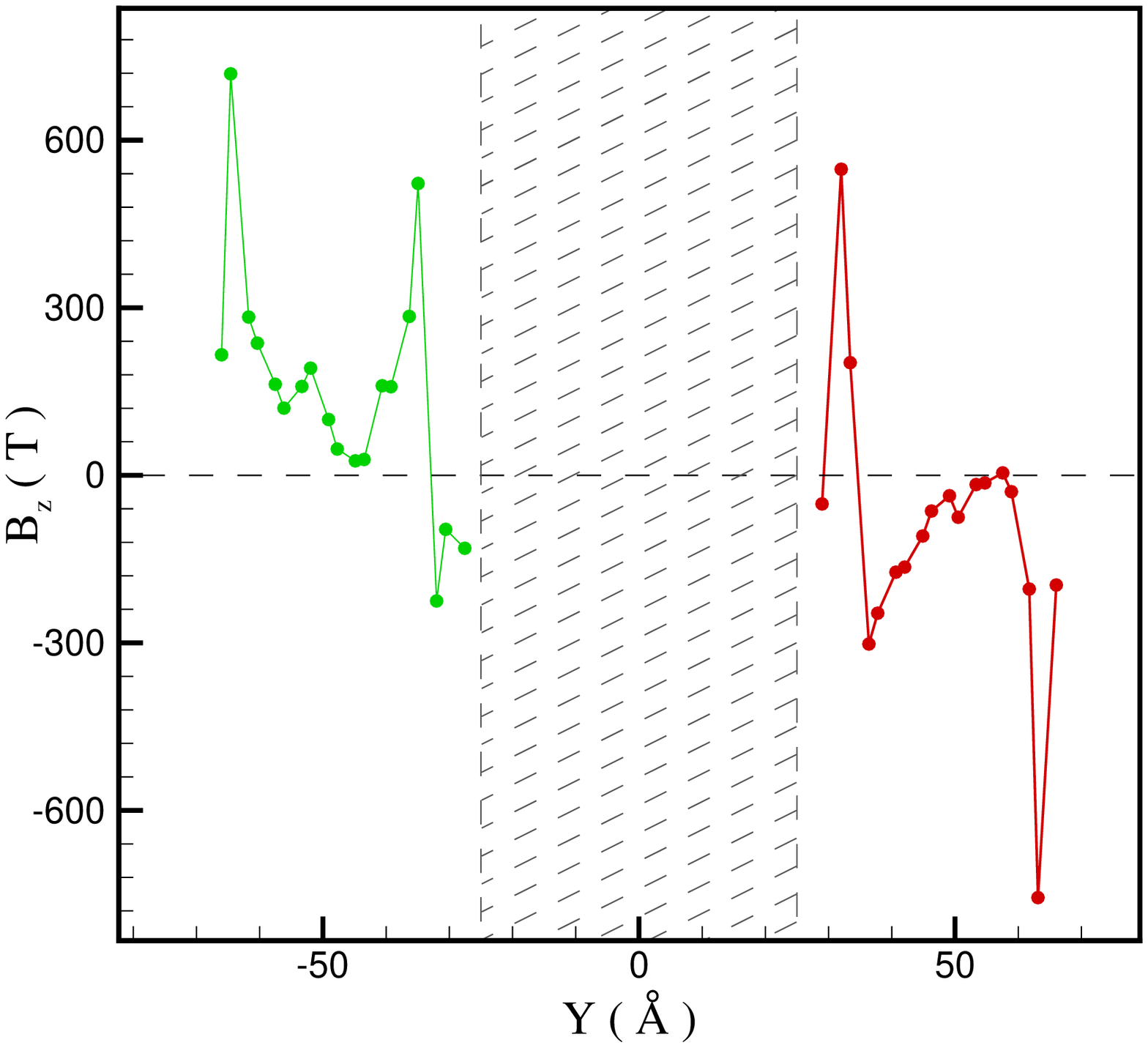}
\caption{(Color online) Pseudo-magnetic field, $B_z$ on a specific
direction indicates in the inset figure, left panel, and along
$y$-direction passing the origin of the ring, right
panel.}\label{bzsd}
\end{center}
\end{figure}

\begin{figure}[h]
\begin{center}
\includegraphics[width=3.3in]{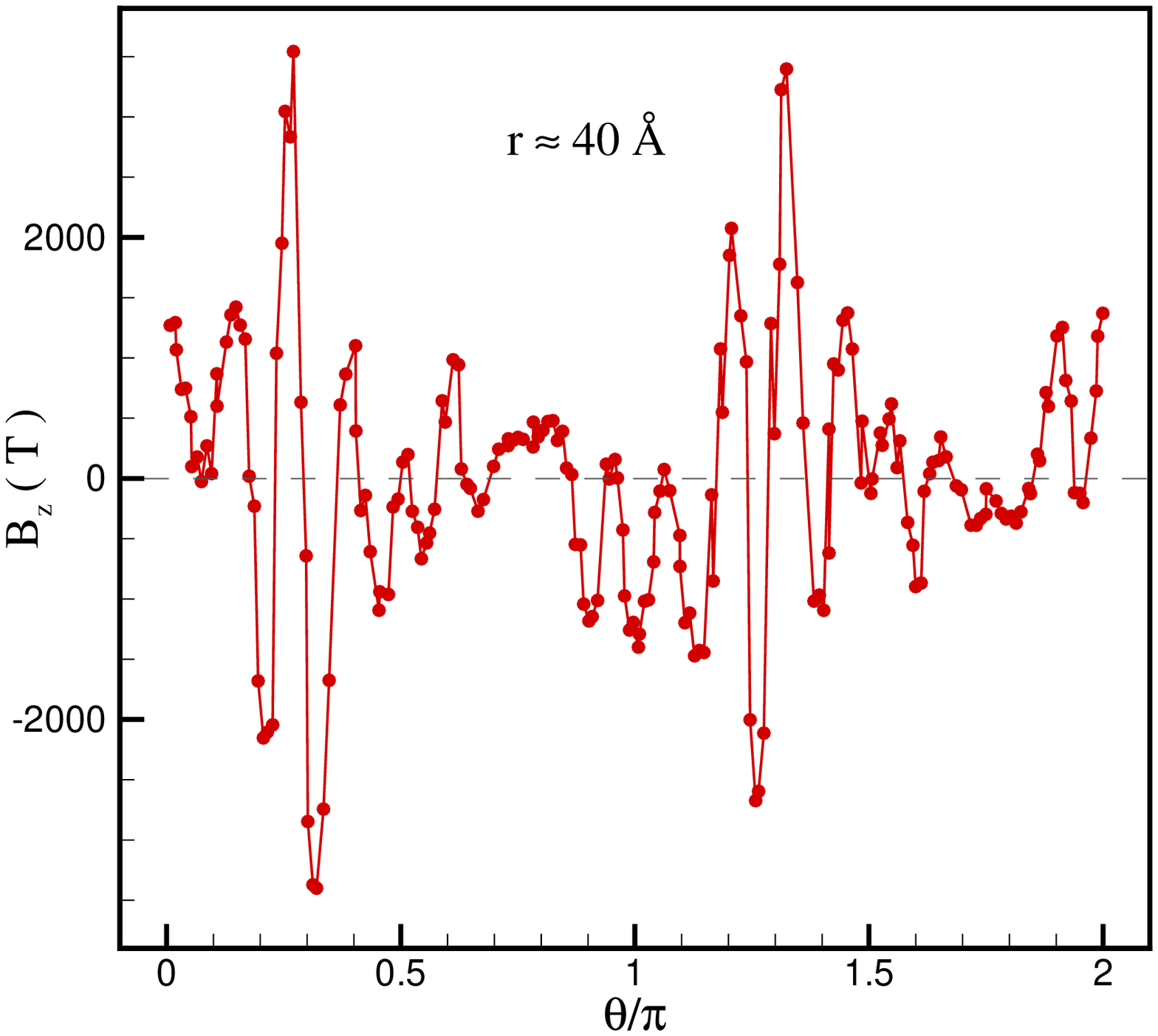}
\includegraphics[width=3.3in]{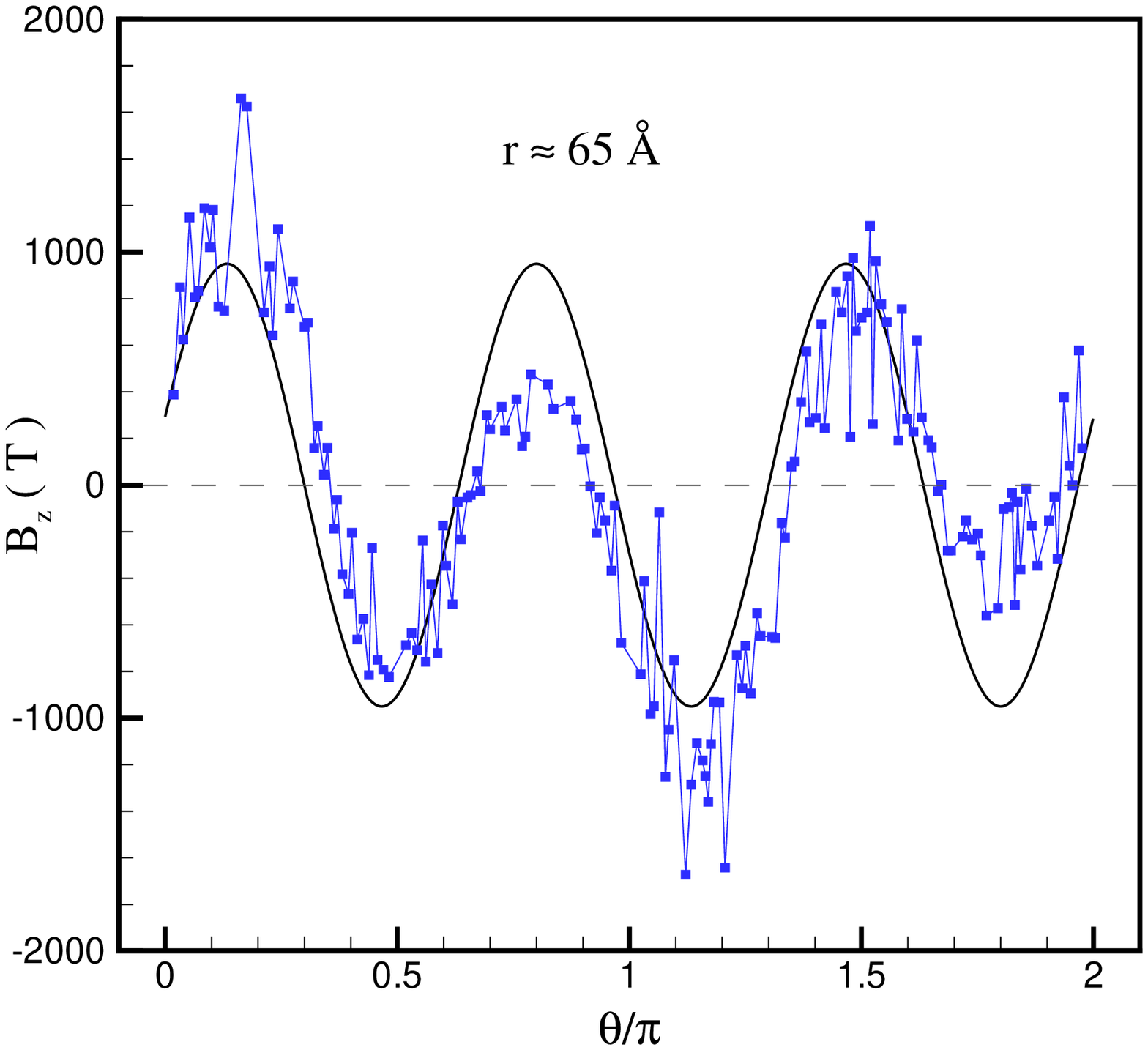}
\caption{(Color online) Pseudo-magnetic field, $B_z$ as a function
of azimuthal angle for case one at $r=4$ nm, left panel and
$r=6.5$nm for case two( right panel). Solid line curve denotes a function which is proportional to $\cos( 3\theta)$.}\label{bzcircle}
\end{center}
\end{figure}

\begin{figure}[h]
\begin{center}
\includegraphics[width=4.in]{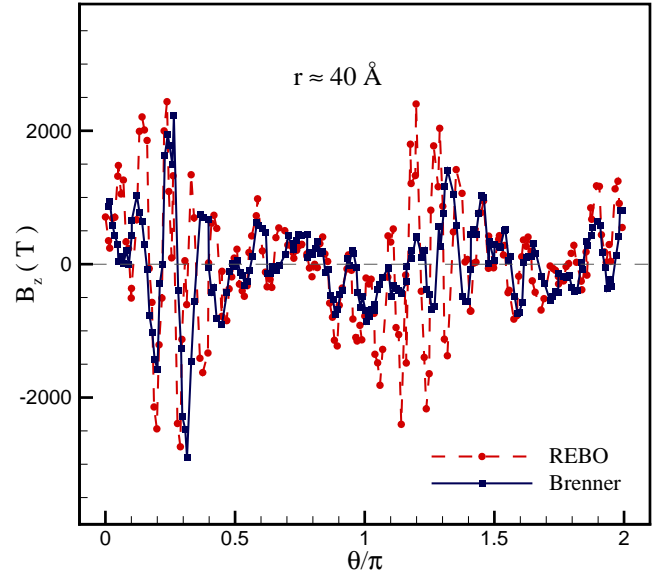}
\caption{(Color online) Pseudo-magnetic field, $B_z$ as a function
of azimuthal angle at $T=50 K$ along a circular ring of radius $4$ nm when the Brenner potential~\cite{brenner} and the REBO potential ~\cite{rebo} are used. Their long-wavelength behavior is very similar to the form $B_z \propto \cos(
3\theta)$, however their short wavelength behavior is different due to the way in which the bond-order is handled in the REBO potential. The outer
boundary of graphene ring, in both cases, is rotated about $\theta_0 = 8^{\circ}$. }\label{bzpot}
\end{center}
\end{figure}

\end{document}